\documentclass[journal,twoside,web]{ieeecolor}
\usepackage{lcsys}
\usepackage{cite}
\usepackage{amsmath,amssymb,amsfonts}
\usepackage{algorithmic}
\usepackage{graphicx}
\usepackage{textcomp}
\usepackage{url}
\usepackage{hyperref}

\def\BibTeX{{\rm B\kern-.05em{\sc i\kern-.025em b}\kern-.08em
    T\kern-.1667em\lower.7ex\hbox{E}\kern-.125emX}}
\begin{document}
\title{Exponentially Convergent Direct Adaptive Pole Placement Control of Plants with Unmatched Uncertainty under FE Condition}
\author{A. Glushchenko, \IEEEmembership{Member, IEEE}, and K. Lastochkin
\thanks{This research was financially supported by Grants Council of the President of the Russian Federation (project MD-1787.2022.4). }
\thanks{Anton Glushchenko is with V.A. Trapeznikov Institute of Control Sciences of RAS, Moscow, Russia (e-mail: aiglush@ipu.ru). }
\thanks{Konstantin Lastochkin is with V.A. Trapeznikov Institute of Control Sciences of RAS, Moscow, Russia (e-mail: lastconst@ipu.ru).}}

\maketitle

\thispagestyle{empty} 

\begin{abstract}
A new method of direct adaptive pole placement control (APPC) is developed for plants with unmatched uncertainty, which linearly depends on a state vector. It guarantees the exponential stability of a control system and exponential convergence of control law adjustable parameters to their true values when the regressor is finitely exciting. Considering the known classical APPC schemes and adaptive methods with exponential regulation, the advantages of the proposed one are that it does not require a priori information on a control input matrix and ensures the monotonic transient behavior of each adjustable parameter of the control law. The theoretical results are supported by the numerical experiments.
\end{abstract}


\section{Introduction}
\label{sec:introduction}
\IEEEPARstart{M}{odel} Reference Adaptive Control (MRAC) is a well-studied and efficient practice-oriented methodology to provide asymptotic stability of plants with significant parameter uncertainty \cite{b1, b2}. \textcolor{black}{Considering a constant setpoint tracking, it is well-known \cite{b3, b4} that the conventional MRAC schemes do not ensure exponential parameter/state convergence to their ground-truth/desired values in the absence of the strict requirement of the regressor {\it{persistent excitation}} (PE).} Generally speaking, the exponential stability is an advantageous property, because it automatically provides fast adaptation and strong self-recovery property, i.e. the uniform ultimate boundedness of the origin tracking and parameter errors in the presence of disturbances.

\textcolor{black}{In this regard, a lot of exponential regulation methods have been proposed in recent years. In \cite{b5} an adaptive law is pre-multiplied with an exponential function to obtain time-varying adaptive gain, which compensates for the regressor excitation vanishing and provides the exponential convergence of the tracking error with independent of the initial conditions and user-assignable rate. In \cite{b6, b7} a least-square-based adaptive law is proposed, which provides boundedness of a time-varying adaptive gain and exponential convergence of the tracking and parameter errors to a compact set under the mild excitation requirements. As thoroughly discussed in \cite{b1, b8}, the high-gain adaptation in \cite{b5, b6, b7} may result in noise amplification and unmodelled dynamics instability, which is dramatically critical for a practical scenario \cite{b9}. In their turn, in contrast to \cite{b5, b6, b7}, some of the methods declared in \cite{b10, b11, b12, b13, b14, b15, b16, b17} avoid high-gain injection and provide exponential rate of convergence with the help of a composite adaptive law. Such methods add some-kind of memory into the conventional adaptive law, which is implemented as a data-stack \cite {b10, b17} or integral-based component \cite{b11, b12, b13, b14, b15, b16} and preserves the exponential rate of convergence when the regressor is finitely (FE) or initially (IE) exciting. The above-considered methods \cite{b5, b6, b7, b10, b11, b12, b13 ,b14, b15, b16, b17} ensure exponential regulation, but do not provide strict guarantee on transient quality of the parameter or tracking errors. To overcome that, in \cite{b18, b19}  adaptive laws are proposed, which, in addition to the exponential regulation properties, provide element-wise monotonicity of the parameter error convergence under FE/IE. It may be considered as a new control quality metric for the adaptive systems \cite{b9}.}

\textcolor{black}{Most of the above-considered exponential regulation methods \cite{b6, b7, b10, b11, b12, b13, b14, b15, b16, b17, b18, b19} still require the plant model to satisfy some structural assumptions named Erzberger’s matching conditions \cite{b20}. It means that the uncertainty must fit completely into the space spanned by the control input matrix. It is a general assumption and starting point for the above-mentioned studies, but many practical scenarios do not follow it \cite{b21}.}

When the Erzberger’s conditions are violated, methods of the {\it adaptive backstepping} (ABS) \cite{b5, b22} can be applied to solve the adaptive control problems. \textcolor{black}{Their disadvantages include a high dynamic order of both the control and adaptive laws.} In order to solve the adaptive control problem for the plants with unmatched uncertainty and avoid the laws of high order, in \cite{b23, b24, b25} it is proposed to identify the uncertainty, which could not be compensated. Then, using the {\it linear matrix inequalities} (LMI), a reference model is derived, which is robust to such uncertainty. \textcolor{black}{Compared to ABS, in the above-mentioned studies the reference model could not be chosen arbitrarily.} Another approach to derive adaptive control laws in presence of the unmatched uncertainty is the {\it adaptive pole placement control} (APPC) schemes \cite{b1, b2}. When the uncertainty depends linearly on the plant states (plants are in a strict-feedback form \cite{b22}), it allows one to provide the required control quality by assigning poles of the closed-loop system. \textcolor{black}{According to \cite{b1, b26}, indirect APPC schemes have gained a lot of attention of researchers over several previous decades, e.g. \cite{b27}. The direct APPC designs \cite{b28} are not so popular, despite the fact that they do not face the control equation solvability issue. The reason is that such schemes require a priori information on the control input matrix, and exponential convergence of the adjustable parameters/tracking error is guaranteed only if the PE condition holds. But, considering the modern methods of exponential regulation \cite{b18, b19}, another effort is worth making to overcome the stated disadvantages.}

\textcolor{black}{So, now, based on the above literature review, we are ready to formulate the contribution of this study: 1) a new method of continuous-time direct APPC is proposed for the plants with unmatched uncertainty, which linearly depends on a state, 2) it guarantees that the parameter/state values converge exponentially to their ground-truth/desired values under FE, 3) the transients monotonicity of the controller adjustable parameters is provided, 4) the common APPC requirement to know sign/elements of the control input matrix is relaxed. To the best of authors’ knowledge, the mentioned properties are provided simultaneously for the first time.}

The remainder of the paper is organized as follows. Section II contains the formal problem statement, the main result is presented in Section III, the results of the numerical experiments are shown in Section IV.

\subsection{Notation}
Further the following notation is used: \textcolor{black}{ $\mathbb{R}^{n}$ and ${\mathbb{R}} ^{n \times m}$  denote the sets of $n$-dimensional real vectors and ${n \times m}$-dimensional real matrices respectively,} $\lvert{.}\rvert$ represents the absolute value, $\lVert{.}\rVert$ denotes Euclidean norm of a vector, $\lambda_{\min }\left( . \right)$ and $\lambda_{\max }\left( . \right)$ are the matrix minimum and maximum eigenvalues respectively, $\sigma \{ . \}$ is the algebraic spectrum of the matrix eigenvalues, $vec\left( . \right)$ is the matrix vectorization operation when its columns are stacked one under another, $vec^{-1}\left( . \right)$ is the inverse operation to $vec\left( . \right)$, \textcolor{black}{$det\{.\}$ stands for a matrix determinant, $adj\{.\}$ -- for an adjoint matrix, $L_{\infty}$ is the space of all essentially bounded functions, $\otimes$ denotes the matrix Kronecker product. The identity and nullity $n \times m$-dimensional matrices are denoted as ${I_{n \times m}}$ and ${0_{n \times m}}$. We also use the fact that for all (possibly singular) ${n \times n}$ matrices $M$ the following holds: $adj \{M\} M = det \{M\}I_{n \times n}$.}

The following definitions are adopted from \cite{b1, b2}:

\textbf{\emph{Definition 1.}}\emph{ The regressor $\varphi\left(t \right)$ is persistently exciting 
$\left({\varphi\left(t \right)\in{\rm{PE}}}\right)$ if $\forall t \ge 0 \; \exists T > 0$ and $\alpha  > 0$ such that: 
\begin{equation}\label{eq1}
  \begin{gathered} 
\small\int \limits_t^{t + T} {\varphi \left( \tau  \right){\varphi ^{\rm{T}}}\left( \tau  \right)d} \tau  \ge \alpha {I_{n \times n}},
\end{gathered}
\end{equation}
where $\alpha$  is the excitation level.
}

\textbf{\emph{Definition 2.}}\emph{ The regressor $\varphi\left(t \right)$ is finitely exciting 
$\left({\varphi\left(t \right)\in{\rm{FE}}}\right)$ over $\left[{t_r^+; \; {\rm{ }}{t_e}}\right]$ if there exist ${t_e} \ge t_r^ +  > 0$  and $\alpha  > 0$ such that the following holds:
\begin{equation}\label{eq2}
  \begin{gathered} 
\small\int \limits_{t_r^ + }^{{t_e}} {\varphi \left( \tau  \right){\varphi ^{\rm{T}}}\left( \tau  \right)d} \tau  \ge \alpha {I_{n \times n}}.
  \end{gathered}
\end{equation}
}

\textcolor{black}{
Let the corollary of Lemma 3.5.4 from \cite{b1} be introduced.
}

\textcolor{black}{
\textbf{\emph{\textcolor{black}{Corollary 1.}}}\emph{ For any $d > 0$, any Hurwitz matrix $A \in {\mathbb{R}^{n \times n}}$, any vector $B \in {\mathbb{R}^{n}}$ there exists matrix $P = P^{\rm{T}} > 0$, a vector $q \in {\mathbb{R}^{n}}$, and a constant $\mu > 0$ such that:}
\begin{equation}\label{eq3}
  \begin{gathered}
{A^{\rm{T}}}P + PA =  - q{q^{\rm{T}}} - \mu P{\rm{, }}\;\;PB = \sqrt {2d} q.
  \end{gathered}
\end{equation}}

\section{Problem Statement}

Let a control problem of linear plants \textcolor{black}{in the strict-feedback form \cite{b22}} be considered:
\begin{equation}\label{eq4}
\textcolor{black}{\forall t \ge t_{r}^+} \begin{array}{l}
{{\dot x}_1}\left( t \right) = {x_2}\left( t \right) + w_1^{\rm{T}}x\left( t \right){\rm{,}}\\
{{\dot x}_2}\left( t \right) = {x_3}\left( t \right) + w_2^{\rm{T}}x\left( t \right){\rm{,}}\\
 \vdots \\
{{\dot x}_n}\left( t \right) = bu\left( t \right) + w_n^{\rm{T}}x\left( t \right){\rm{, }}\; x\left( \textcolor{black}{t_{r}^+} \right) = {x_0}{\rm{,}}\\
\textcolor{black}{y\left(t\right)=h^{\rm T}x\left(t\right).}
\end{array}
\end{equation}
where $x\left( t \right) \in {\mathbb{R}^n}$ is a state vector with unknown initial values $x_0$ , $u\left( t \right) \in \mathbb{R}$ is a control signal,  \textcolor{black}{$y\left(t\right)$ is the system output}, ${w_i} \in {\mathbb{R}^n}$ is a vector of the unknown \textcolor{black}{time-invariant parameters} $i = \overline {1,n} {\rm{,}}$ $b \in \mathbb{R}$ is an unknown constant gain, \textcolor{black}{$h \in \mathbb{R}^{n}$ is the vector to form $y\left(t\right)$}.

As each $w_i^{\rm{T}}x\left( t \right)$ is linear, the system \eqref{eq4} is rewritten as:
\begin{equation}\label{eq5}
\textcolor{black}{
\begin{gathered}
\left\{ {\begin{array}{l}\dot x\left( t \right) = \theta _{AB}^{\rm{T}}\Phi \left( t \right) = Ax\left( t \right) + Bu\left( t \right){\rm{, }}\;\;x\left( \textcolor{black}{t_{r}^+} \right) = {x_0}{\rm{,}}\\
\textcolor{black}{y\left(t\right)=h^{\rm T}x\left(t\right).}
\end{array}}\right.\\
\Phi \left( t \right) = {{\begin{bmatrix}
{{x^{\rm{T}}}\left( t \right)}&{\textcolor{black}{u\left( t \right)}}
\end{bmatrix}}^{\rm{T}}}{\rm{, }}\;\;\theta _{AB}^{\rm{T}} = {\begin{bmatrix}
A&B
\end{bmatrix}}{\rm{,}}
\end{gathered}}
\end{equation}
where $A \in {\mathbb{R}^{n \times n}}$ is an unknown state matrix, which $i^{\rm th}$ row corresponds to ${x_{i + 1}}\left( t \right) + w_i^{\rm{T}}x\left( t \right)$, and $n^{\rm th}$ – to $w_n^{\rm{T}}x\left( t \right)$ respectively, $B \in {\mathbb{R}^n}$ is an unknown control input vector, which $n^{\rm th}$ row contains the parameter $b$. It is assumed that the \textcolor{black}{pairs $\left(A, B\right)$ and $\left(A, h^{\rm T}\right)$ are controllable and observable respectively}, $\Phi \left( t \right) \in {\mathbb{R}^{n + 1}}$ is measurable $\forall t > \textcolor{black}{t_{r}^+},\;$ ${\theta _{AB}} \in {\mathbb{R}^{\left( {n + 1} \right) \times n}}$ is time-invariant and unknown.

The required transients quality for the system output $y\left(t\right)$ is defined as a modal model (generator) \cite{b29}:
\begin{equation}\label{eq6}
\left\{ \begin{array}{l}
\dot \chi \left( t \right) = \Gamma \chi \left( t \right){\rm{, }}\\
v\left( t \right) = {h^{\rm{T}}}\chi \left( t \right){\rm{,}}
\end{array} \right.\chi \left( \textcolor{black}{t_{r}^+} \right) = {\chi _0}{\rm{,}}
\end{equation}
where $\chi \left( t \right) \in {\mathbb{R}^n}$ is a generator state with initial conditions ${\chi _0}$, $v\left( t \right) \in \mathbb{R}$ is a generator output, $\Gamma  \in {\mathbb{R}^{n \times n}}$ is a Hurwitz state matrix with desired poles placement, the pair $(\Gamma, h^{\rm{T}})$ is assumed to be observable.

The following proposition holds for \eqref{eq5}, which required control quality for $y\left(t\right)$ is defined as the modal model \eqref{eq6}:

\textbf{\emph{Proposition 1.}}\emph{  Let $\sigma \left\{ A \right\}\, \cap \, \sigma \left\{ \Gamma  \right\} = \textcolor{black}{\emptyset}$, then, considering that the pairs $(A, B)$ and $(\Gamma, h^{\rm T})$ are controllable and observable respectively, the first equation of \eqref{eq5} is rewritten as: 
\begin{equation}\label{eq7}
  \begin{gathered} 
   \dot x\left( t \right) = {A_\Sigma }x\left( t \right) + B\left( {u\left( t \right) - {K_x}x\left( t \right)} \right){\rm{,}}
  \end{gathered}
\end{equation}
where ${A_\Sigma } = A + B{K_x}{\rm{, }}\;\;\sigma \left\{ {{A_\Sigma }} \right\} = \sigma \left\{ \Gamma  \right\}$, and $K_x^{\rm{T}} \in {\mathbb{R}^n}$ is defined as a solution of the set of equations:
\begin{equation}\label{eq8}
  \begin{gathered} 
\left\{ \begin{array}{l}
M\Gamma  - AM = B{h^{\rm{T}}}{\rm{,}}\\
{h^{\rm{T}}} = {K_x}M{\rm{,}}
\end{array} \right.
  \end{gathered}
\end{equation}
where $M \in {\mathbb{R}^{n \times n}}$ is a nonsingular $\left( {\exists {M^{ - 1}}} \right)$ matrix of a linear conform transformation $x\left( t \right) = M\chi \left( t \right)$.}

\emph{Proofs of the necessity and sufficiency of the following requirements: 1) $(A,\; B)$ controllability, 2) $(\Gamma, \; h^{\rm T})$ observability, 3) \textcolor{black}{$\sigma \left\{ A \right\} \cap \sigma \left\{ \Gamma  \right\} = \emptyset$} are shown in} \cite{b29, b30}.

Let the required control quality of \eqref{eq7} output for the steady-state and transient modes be defined as the reference model:
\begin{equation}\label{eq9}
\begin{gathered} 
{\dot x_{ref}}\left( t \right) = {A_\Sigma }{x_{ref}}\left( t \right) + {B_{ref}}r\left( t \right){\rm{,}}\;{x_{ref}}\left( \textcolor{black}{t_{r}^+} \right) = {x_{0ref}}{\rm{,}}
\end{gathered}
\end{equation}
where ${x_{ref}}\left( t \right) \in {\mathbb{R}^n}$ is an unmeasurable state of the reference model with the initial values ${x_{0ref}}$, ${B_{ref}} = B{K_r} \in {\mathbb{R}^n}$ is an unknown input vector of the reference model, ${K_r} \in \mathbb{R}$ is an unknown parameter, $r\left( t \right) \in \mathbb{R}$ \textcolor{black}{denotes a bounded, piecewise continuous reference input signal}.

The parameter ${K_r}$ is defined from the fact that the reference model output  ${y_{ref}}\left( t \right) = {h^{\rm{T}}}{x_{ref}}\left( t \right)$ should be equal \textcolor{black}{to the  $r\left( t \right)$ at the steady-state mode if $r\left( t \right)$ is constant:}
\begin{equation}\label{eq10}
  \begin{gathered} 
\begin{array}{c}
{K_r} = \arg \left\{ {h^{\rm{T}}}{{\left( {sI - {A_\Sigma }} \right)}^{ - 1}}\left| {_{s = 0}} \right.{B_{ref}} = \right.\\
\left. =  - {h^{\rm{T}}}A_\Sigma ^{ - 1}B{K_r} = 1 \right\} =  - {\left( {{h^{\rm{T}}}A_\Sigma ^{ - 1}B} \right)^{ - 1}.}
\end{array}
  \end{gathered}
\end{equation}

So the error equation between \eqref{eq7} and \eqref{eq9} is written as:
\begin{equation}\label{eq11}
\begin{array}{c}
{{\dot e}_{ref}}\left( t \right) = {A_\Sigma }{e_{ref}}\left( t \right) + B\left( {u\left( t \right) - {\theta ^{\rm{T}}}\left( t \right)\omega \left( t \right)} \right){\rm{,}}\\
\textcolor{black}{\omega \left( t \right) = { {\begin{bmatrix}
{{x^{\rm{T}}}\left( t \right)}&{r\left( t \right)}
\end{bmatrix}}^{\rm{T}}},} \; \textcolor{black}{{\rm{ }}{\theta ^{\rm{T}}}{\rm{ = }} {\begin{bmatrix}
{{K_x}}&{{K_r}}
\end{bmatrix}}.}
\end{array}
\end{equation}
where ${e_{ref}}\left( t \right) = x\left( t \right) - {x_{ref}}\left( t \right)$ is an unmeasurable tracking error, $\omega \left( t \right) \in {\mathbb{R}^{n + 1}}$ is a regressor.

Using \eqref{eq11}, the control law $u\left( t \right)$ is defined as:
\begin{equation}\label{eq12}
u\left( t \right) = \textcolor{black}{{\hat \theta ^{\rm{T}}}\left( t \right)\omega \left( t \right)} = {\hat K_x}\left( t \right)x\left( t \right) + {\hat K_r}\left( t \right)r\left( t \right){\rm{,}}
\end{equation}
where $\hat K_x^{\rm{T}}\left( t \right) \in {\mathbb{R}^n}{\rm{, }}\;{\hat K_r}\left( t \right) \in \mathbb{R}$ are the adjustable parameters and it is assumed that ${\hat K_r}\left( 0 \right) \ne 0$.

Substituting \eqref{eq12} into \eqref{eq11}, it is obtained:
\begin{equation}\label{eq13}
{\dot e_{ref}}\left( t \right) = {A_\Sigma }{e_{ref}}\left( t \right) + B{\tilde \theta ^{\rm{T}}}\left( t \right)\omega \left( t \right){\rm{,}}
\end{equation}
where $\tilde \theta \left( t \right) \in {\mathbb{R}^{n + 1}}$ is the parameter error. 

Using \eqref{eq13}, let the main goal of the study be formulated.

\textbf{\emph{Goal.}} \emph{ \textcolor{black}{We consider the system \eqref{eq4} with unknown $A,\: B$ and the desired observable pair $(\Gamma, h)$. Let $\theta$ be defined as in \eqref{eq11}, where $K_x$ gives the desired pole placement and $K_r$ is defined in \eqref{eq10}, such values are unique. Then the goal is to derive an update law for $\hat\theta\left(t\right)$ such that, when $\Phi \left( t \right) \in {\rm{FE}}$, the control law \eqref{eq12} stabilizes the system, providing the desired pole placement, and ensures exponential convergence:}
\begin{equation}\label{eq14}
\mathop {\lim }\limits_{t \to \infty } \left\| {\xi \left( t \right)} \right\| = 0 \;{\rm{ (exp)}}{\rm{,}}
\end{equation}
where $\xi \left( t \right) = [{\begin{smallmatrix}
{e_{ref}^{\rm{T}}\left( t \right)}&{{{\tilde \theta }^{\rm{T}}} \left( t \right)} \end{smallmatrix}]^{\rm{T}}}$ is augmented error.}

\textbf{\emph{Remark 1.}}\emph{ The modal model \eqref{eq6} defines the algebraic spectrum of the matrix ${A_\Sigma }$, and the required transient behavior of the plant \eqref{eq7} with an accuracy up to the linear transformation $x\left( t \right) = M\chi \left( t \right)$. At the same time, the reference model \eqref{eq9} defines both the steady-state and transient behavior of the plant \eqref{eq7}. However, unlike \eqref{eq6}, it could not be implemented as ${A_\Sigma }{\rm{, }}\;{B_{ref}}$ are unknown.}

\section{Main Result}
The proposed solution of the problem \eqref{eq14} is based on a new two-stage procedure of the inverse parameterization of the direct adaptive control problem, which consists of the following steps. \textcolor{black}{\textbf{Step 1.} Using the DREM procedure \cite{b31,b32}, the regression equations with a scalar regressor with respect to the unknown matrices $A,\;B$ are derived. \textbf{Step 2.} Using analytical expressions \eqref{eq8} and \eqref{eq10}, the obtained equations are transformed into the regressions with respect to the unknown control law parameters $\theta$.}

\textcolor{black}{As for Step 1 and Step 2, the system matrices $\left( A \right.$ and $\left. B \right)$ and the state derivative $\dot x\left( t \right)$ are considered to be unknown, and only measured state $x\left( t \right)$ along the system trajectory and the control input $u \left( t \right)$ will be used for the proposed adaptive control design.}

\textcolor{black}{\textbf{Step 1.} To achieve the goal \eqref{eq14}, first of all, according to \cite{b18} (Lemma 1) and \cite{b19}, we introduce filtering dynamics for \eqref{eq5} in the linear regression form:
\begin{equation}\label{eq15}
\begin{array}{c}
\overline z\left( t \right) = x\left( t \right) - l\overline x\left( t \right)  = \overline \theta _{AB}^{\rm{T}}\overline \varphi \left( t \right){\rm{,}}\\
\overline \varphi \left( t \right) = {{\begin{bmatrix}
{{{\overline \Phi }^{\rm{T}}}\left( t \right)}&{{e^{ - lt}}}
\end{bmatrix}} ^{\rm{T}}}{\rm{, }}\; \overline \theta _{AB}^{\rm{T}} = {\begin{bmatrix}
A&B&{x\left( \textcolor{black}{t_{r}^+} \right)}
\end{bmatrix}} {\rm{,}}\\
\dot {\overline \Phi} \left( t \right) =  - l\overline \Phi \left( t \right) + \Phi \left( t \right){\rm{, }}\;\overline \Phi \left( \textcolor{black}{t_{r}^+} \right) = {0_{n + 1}},
\end{array}
\end{equation}
where $l > 0$, $\overline z\left( t \right)$  is the measurable function, $\overline \varphi \left( t \right) \in {\mathbb{R}^{n + 2}}$ is the measurable regressor, $\overline \theta _{AB}^{\rm{T}} \in {\mathbb{R}^{n \times \left( {n + 2} \right)}}$ is the augmented vector of the unknown parameters, $\overline x\left( t \right)\in \mathbb{R}^{n }$ is the first $n$ elements of the vector $\overline \Phi \left( t \right)\in \mathbb{R}^{n + 1}$. More details on how to obtain \eqref{eq15} from \eqref{eq5} could be found in \cite{b33} (Section I).}

\textcolor{black}{Considering the structure and excitation properties of the regressor $\overline \varphi \left( t \right)$, the following assumption is introduced.}

\textcolor{black}{\textbf{\emph{Assumption 1.}}\emph{ The parameter $l$ in \eqref{eq15} is chosen so as the implication $\overline \Phi \left( t \right) \in {\rm{FE}} \Rightarrow \overline \varphi \left( t \right) \in {\rm{FE}}$ holds.}}

\textcolor{black}{Applying the DREM procedure \cite{b31, b32}, the equation \eqref{eq15} with the vector regressor $\overline \varphi \left( t \right)$ is transformed into the one with the scalar regressor as follows:
\begin{equation}\label{eq16}
\begin{array}{c}
z\left( t \right) = \varphi \left( t \right){{\overline \theta }_{AB}}{\rm{,}}\\
z\left( t \right){\rm{:}} = adj\left\{ {\mathfrak{H}\left[ {\overline \varphi \left( t \right){{\overline \varphi }^{\rm{T}}}\left( t \right)} \right]} \right\}\mathfrak{H}\left[ {\overline \varphi \left( t \right){{\overline z}^{\rm{T}}}\left( t \right)} \right],\\
\varphi \left( t \right){\rm{:}} = det \left\{ {\mathfrak{H}\left[ {\overline \varphi \left( t \right){{\overline \varphi }^{\rm{T}}}\left( t \right)} \right]} \right\}{\rm{,}}
\end{array}
\end{equation}
where $\mathfrak{H}\left[ . \right]{\rm{:}} = {1 \mathord{\left/ \right.} {\left( {p + k} \right)}}\left[ .\right]$ is a linear stable operator with $k > 0$ and $p:= {d \over dt}$. $z\left( t \right) \in {\mathbb{R}^{\left( {n + 2} \right) \times n}}$ and $\varphi \left( t \right) \in \mathbb{R}$ are measurable.}

Considering the definitions of ${\overline \theta _{AB}}$ from \eqref{eq15} and $\varphi \left( t \right) \in \mathbb{R}$ from \eqref{eq16}, the following regression equations are obtained:
\begin{equation}\label{eq17}
\begin{array}{c}
{z_A}\left( t \right) = {z^{\rm{T}}}\left( t \right)  \small{\mathfrak{L} = \varphi \left( t \right)A{\rm{,  }} \;\mathfrak{L} = { {\begin{bmatrix}
{{I_{n \times n}}}&{{0_{n \times 2}}}
\end{bmatrix}} ^{\rm{T}}} {\rm{, }}} \\
{z_B}\left( t \right) = {z^{\rm{T}}}\left( t \right)\mathfrak{e} =\varphi \left( t \right)B{\rm{,} \; \small{\mathfrak{e} = \left[ {{0_{1 \times \it{n}}}}\;1\;\;0\right]^{\rm{T}}}},
\end{array}
\end{equation}
where ${z_A}\left( t \right) \in {\mathbb{R}^{n \times n}}$, ${z_B}\left( t \right) \in {\mathbb{R}^n}$, $\mathfrak{L}$ and $\mathfrak{e}$ have appropriate dimensions.

\textbf{\emph{Remark 2.}}\emph{ According to the results of Lemma 6.8 \cite{b2}, as the filter $1/\left(p+l\right)$ from \eqref{eq15} is stable, if $\Phi \left( t \right) \in {\rm{FE}}$, then $\overline \Phi \left( t \right) \in {\rm{FE}}$. So, following Assumption 1, this results in $\overline \varphi \left( t \right) \in {\rm{FE}}$. In \cite{b32} the implication $\overline \varphi \left( t \right) \in {\rm{FE}} \Rightarrow \varphi \left( t \right) \in {\rm{FE}}$ is proved for the DREM procedure \cite{b16}}.

\textcolor{black}{\textbf{Step 2.} Then, as $\varphi \left( t \right) \in \mathbb{R}$, the aim is to obtain the regression with respect to $\theta$.}

\textbf{\emph{Proposition 2.}}\emph{ Applying the measurable signals ${z_A}\left( t \right){\rm{, }}\;{z_B}\left( t \right){\rm{, }}\;\varphi \left( t \right)$ and the equations \eqref{eq8} and \eqref{eq10}, the following linear regression equation is obtained:
\begin{equation}\label{eq18}
Y\left( t \right) = \Delta \left( t \right)\theta {\rm{,}}
\end{equation}
where $\theta$ is defined in \eqref{eq11}, $Y\left( t \right) \in {\mathbb{R}^{n + 1}}$ and $\Delta \left( t \right) \in \mathbb{R}$ are measurable because ${z_A}\left( t \right){\rm{, }}\;{z_B}\left( t \right){\rm{, }}\;\varphi \left( t \right)$ are measurable, and the implication $\Delta \left( t \right) \in {\rm{FE}} \Rightarrow \Phi \left( t \right) \in {\rm{FE}}$ holds.}

\emph{Proof of Proposition 2 and the definitions of $Y(t)$ and $\Delta(t)$ {\rm (}see \eqref{eqA13}{\rm)} are postponed to Appendix.}

\textcolor{black}{Using \eqref{eq18}, we are in position to introduce the direct law to estimate $\theta$:
\begin{equation}\label{eq19}
\begin{array}{c}
\dot {\hat \theta} \left( t \right) =  - \gamma \Delta \left( t \right)\left( {\Delta \left( t \right)\hat \theta \left( t \right) - Y\left( t \right)} \right),
\end{array}
\end{equation}
where $\gamma > 0$ is the adaptive gain.}

However, based on proved in \cite{b31}, the law \eqref{eq19} provides $\mathop {\lim }\limits_{t \to \infty } \left\| {\tilde \theta \left( t \right)} \right\| = 0\;{\rm{ (exp)}}$ only if $\Delta \left( t \right) \in {\rm{PE}}$. Therefore, the law \eqref{eq19} does not allow one to achieve the goal \eqref{eq14}, and the regression \eqref{eq18} requires additional manipulations. To this end, using the results of \cite{b18}, the regression equation \eqref{eq18} is passed through the filter with exponential forgetting:
\begin{equation}\label{eq20}
\begin{array}{c}
\Upsilon \left( t \right) : =\int\limits_{t_r^ + }^t {{e^{ - \sigma \tau }}\Delta \left( \tau  \right)Y\left( \tau  \right)d\tau } =\Omega \left( t \right)\theta {\rm{,}}\\
\Omega \left( t \right) = \int\limits_{t_r^ + }^t {{e^{ - \sigma \tau }}{\Delta ^2}\left( \tau  \right)d\tau } {\rm{, }} 
\end{array}
\end{equation}
where $\sigma  > 0$, $\Upsilon \left( t \right) \in {\mathbb{R}^{n + 1}}$, $\Omega \left( t \right) \in \mathbb{R}$.

In \cite{b19, b34} it is proved for the regressor $\Omega \left( t \right)$ that:

\textbf{\emph{Proposition 3.}}\emph{  If $\Delta \left( t \right) \in {\rm{FE}}$ over $\left[ {t_r^ + {\rm{; }}{t_e}} \right]$ and $\left( \Delta \left( t \right) \in {L_\infty } \right.$ or  $\left| {\Delta \left( t \right)} \right| \le {c_1}{e^{{c_2}t}}$ and $\left. \sigma  > {\rm{2}}c_2 \right)$, where $c_1 > 0$, $c_2 > 0$, then
\begin{enumerate}
\item $\forall t \ge t_{r}^+ \; \Omega \left( t \right)  \in {L_\infty }{\rm{, }}\;\Omega \left( t \right) \ge 0,$
\item $\forall t \ge {t_e}\;\;{\rm{ }}\Omega \left( t \right) > 0,\;{\rm{ }}{\Omega _{LB}} \le \Omega \left( t \right) \le {\Omega _{UB}}.$
\end{enumerate}}

\emph{Proof of Proposition 3 is shown in} \textcolor{black}{\cite{b19} (Proposition 3), \cite{b34} (Assertion 3)}.

Using \eqref{eq20} and the properties of  $\Omega \left( t \right)$, the following adaptive law is introduced:
\begin{equation}\label{eq21}
\begin{array}{c}
\dot {\hat \theta} \left( t \right) =  - \gamma \Omega \left( t \right)\left( {\Omega \left( t \right)\hat \theta \left( t \right) - \Upsilon \left( t \right)} \right).
\end{array}
\end{equation}

The conditions, when \eqref{eq21} provides the achievement of \eqref{eq14}, is formulated \textcolor{black}{on the basis of results in \cite{b19}} as \emph{Theorem}.

\textbf{\emph{Theorem 1.}}\emph{  Let $\Phi \left( t \right) \in {\rm{FE}}$, then, if the parameter $\gamma$ is chosen in accordance with
\begin{equation}\label{eq22}
\begin{array}{c}
\gamma  = \left\{ \begin{array}{l}
{\rm{0}}{\rm{, \;if\; }}\Omega \left( t \right) = 0,\\
{\textstyle{{{\gamma _0}{\lambda _{{\rm{max}}}}\left( {\omega \left( t \right){\omega ^{\rm{T}}}\left( t \right)} \right) + {\gamma _1}} \over {{\Omega ^2}\left( t \right)}}}\;{\rm{ otherwise}}{\rm{,}}
\end{array} \right.
\end{array}
\end{equation}
then the adaptive law \eqref{eq21} provides the following properties:
\begin{enumerate}
\item $\forall {t_a} \ge {t_b}\;{\rm{ }}\left| {{{\tilde \theta }_i}\left( {{t_a}} \right)} \right| \le \left| {{{\tilde \theta }_i}\left( {{t_b}} \right)} \right|{\rm{;}}$
\item $\forall t \ge t_{r}^+ \; \xi \left( t \right) \in {L_\infty }{\rm{;}}$
\item $\forall t \ge {t_e}$ the error $\xi \left( t \right)$ converges exponentially to zero at the rate, which minimum value is directly proportional to the parameters ${\gamma _0} \ge 1$ and ${\gamma _1} \ge 0$.
\end{enumerate}}

\textcolor{black}{\textcolor{black}{\emph{Proof of Theorem can be found in} \cite{b19} {\it (Theorem)}.}}

Thus, the proposed system \eqref{eq12}, \eqref{eq21}, \eqref{eq22} provides achievement of the goal \eqref{eq14} and does not require {\it a priori} information on the matrix  $B$, guarantees transient monotonicity of each adjustable parameter of the control law \eqref{eq12}. It also allows one to solve problem of the adaptive control of the plants with unmatched linear uncertainty \eqref{eq4} without application of ABS \cite{b22} or classical APPC \cite{b1,b26,b27,b28} schemes, which are complex and difficult for practical implementation.

\textcolor{black}{\textbf{\emph{Remark 3.}}\emph{ The proposed adaptive law \eqref{eq21} provides asymptotic stability of ${e_{ref}} \left( t \right)$ with exponential rate of convergence only when $\Phi \left( t \right) \in {\rm{FE}}$. Therefore, to implement \eqref{eq21} in practice, the {\it a priori} information is required that this condition holds. So, future scope of our research is two-fold. First of all, the conditions on $r\left( t \right)$ and $\hat \theta \left( \textcolor{black}{t_{r}^+} \right)$ will be obtained, under which the requirement $\Phi \left( t \right) \in {\rm{FE}}$ is met for the whole class of the linear strict-feedback plants \eqref{eq4}. Secondly, modifying the parametrization \eqref{eq15}, \eqref{eq16}, \eqref{eq17}, the adaptive law will be derived, which ensures exponential regulation under strictly weaker semi-${\rm FE}$ condition.}}

\section{Numerical Experiment}
A numerical experiment was conducted in Matlab/Simulink. The plant \eqref{eq4} and modal model \eqref{eq6} were chosen as follows:
\begin{equation}\label{eq23}
\left\{ {\begin{array}{*{20}{ll}}
\begin{array}{ll}
\dot x\left( t \right)\! =\! {\begin{bmatrix}5&{ - 2}\\
4&2
\end{bmatrix}} x\left( t \right)\! +\! {\begin{bmatrix} 0\\
2
\end{bmatrix}} u\left( t \right){\rm{, }}\;x\left( 0 \right) \!=\! {\begin{bmatrix}
0\\
{ - 1}
\end{bmatrix}} {\rm{,}}\\
\textcolor{black}{y\left(t\right)=\begin{bmatrix}1&0\end{bmatrix}x\left(t\right),}
\end{array}
\end{array}} \right.
\end{equation}

\begin{equation}\label{eq24}
\begin{array}{c}
\left\{ {\begin{array}{*{20}{ll}}
{\dot \chi \left( t \right) = {\begin{bmatrix}
{ - 4}&1\\
{ - 8}&0
\end{bmatrix}} \chi \left( t \right){\rm{,}}}\\
{v\left( t \right) = {\begin{bmatrix}
1&0
\end{bmatrix}} \chi\left( t \right) {\rm{,}}}
\end{array}} \right.\chi \left( 0 \right) = {\begin{bmatrix}
0\\
0
\end{bmatrix}}.
\end{array}
\end{equation}

\textcolor{black}{According to the problem statement, all plant \eqref{eq23} parameters and initial conditions were considered as unknown.} The values of the setpoint $r\left(t\right),\; \gamma_0,\; \gamma_1, $ the parameters of \eqref{eq15}, \eqref{eq16}, \eqref{eq20} and the initial values of the adjustable parameters of the control law \eqref{eq12} were set as follows:
\begin{equation}\label{eq25}
\begin{array}{c}
r = \!l = \!{\gamma _0}\! = \!1{\rm{, }}\:k\! =\! 10{\rm{,}}\:\sigma \! =\! {\textstyle{5 \over 1}}\:{\rm{,}} {\gamma _1}\! =\! 0{\rm{,}}\:\hat \theta \left( 0 \right)\! =\! { {\begin{bmatrix}
0\!\!\!\!&0\!\!\!\!&2
\end{bmatrix}} ^{\rm{T}}}
\end{array}
\end{equation}

Figure 1 shows the transient curves of: 1) the plant \eqref{eq4} and the reference model \eqref{eq9} states when ${x_0} = {x_{0ref}}$, and \textcolor{black}{2) the adjustable parameters $\hat \theta \left( t \right)$ of the control law \eqref{eq12} and their ideal values $\theta$, which were calculated using \eqref{eq8} and \eqref{eq10}.}
   \begin{figure}[thpb]
      \centering
      \includegraphics[scale=0.71]{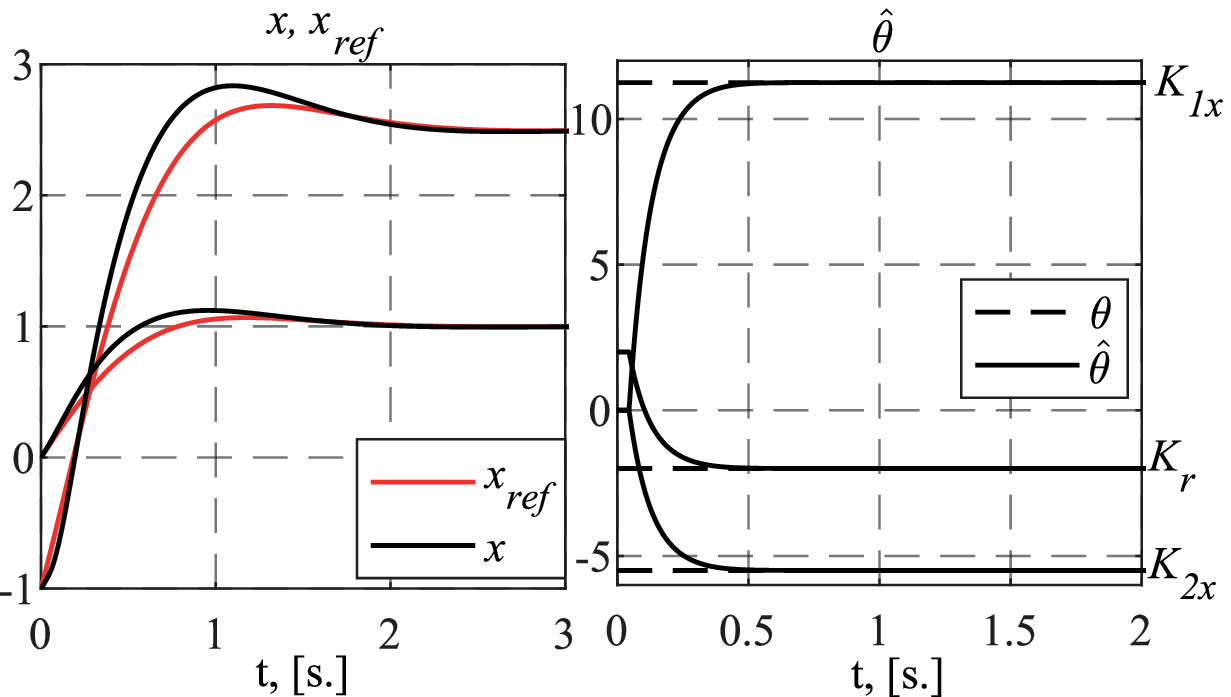}
      \caption{Transients of $x\left( t \right)$, ${x_{ref}}\left( t \right)$ and $\hat \theta \left( t \right)$.}
      \label{Figure1} 
      \end{figure}

Figure 2 shows the transient curves of: 1) the control signal \eqref{eq12} and its ideal value ${u^*}\left( t \right)$, which was calculated on the basis if the ground-truth values of the controller parameters, and 2) the regressor $\Omega \left( t \right)$  obtained from \eqref{eq20}.
   \begin{figure}[thpb]
      \centering
      \includegraphics[scale=0.95]{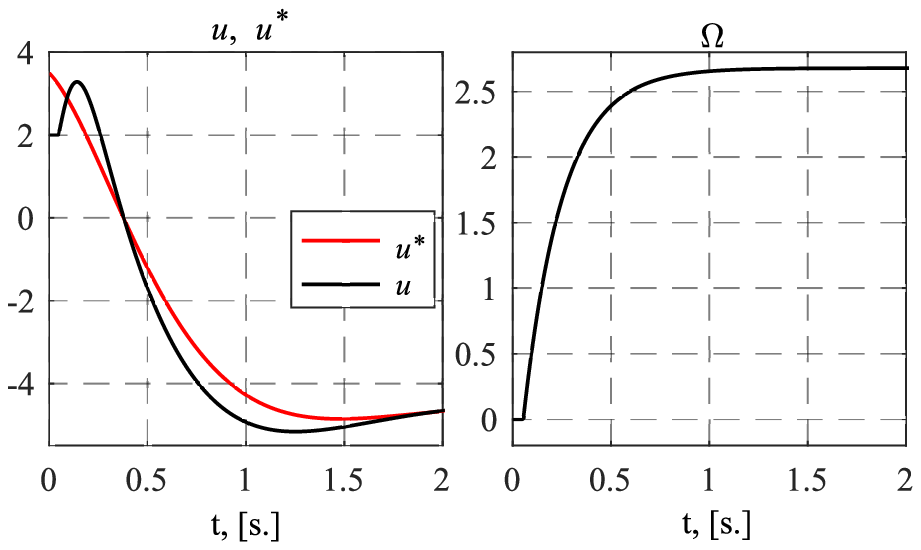}
      \caption{Transients of $u\left( t \right)$, ${u^*}\left( t \right)$ and $\Omega \left( t \right)$.}
      \label{Figure2}
      \end{figure}

The simulation results, which are presented in Fig. 1 and Fig.2, validated the theoretical conclusions made in Proposition 2 and Theorem 1. The developed system solved the stated problem of the direct adaptive pole placement control with exponential rate of convergence \eqref{eq14} and provided elementwise monotonicity of transients of the control law \eqref{eq12} parameters.

\section{Conclusion}
\textcolor{black}{
A new method of the direct adaptive pole placement to control plants with unmatched uncertainty, which linearly depended on the state vector, was developed.}

The proposed two-step procedure of the inverse parameterization of the direct adaptive control problem demonstrated its effectiveness and could be further applied to derive the adaptive laws for various control laws with the known analytical dependence of the ideal parameters on the plant matrices.


\appendices

\renewcommand{\theequation}{A\arabic{equation}}
\setcounter{equation}{0}  

\section*{Appendix}
{\it Proof of Proposition 2.} \textcolor{black}{To prove the proposition, the first aim is to obtain the regression with respect to $K_{x}$.} So, the first equation of \eqref{eq8} is multiplied by $\varphi \left( t \right)$. Then \eqref{eq17} is substituted into the result, \textcolor{black}{and the obtained equation is rewritten using the known vectorization operation properties:}
\begin{equation}
\begin{gathered}
{{\overline Y}_M}\left( t \right){\rm{:}} = vec\left( {{z_B}\left( t \right){h^{\rm{T}}}} \right) = {{\overline \Delta }_M}\left( t \right)vec\left( M \right){\rm{,}}\\
{{\overline \Delta }_M}\left( t \right){\rm{:}} =  - {I_{n \times n}} \otimes {z_A}\left( t \right) + \varphi\left( t \right) {\Gamma ^{\rm{T}}} \otimes {I_{n \times n}} ,
\end{gathered}\normalsize{\label{eqA1}}
\end{equation}
where ${\overline Y_M}\left( t \right) \in \mathbb{R}^{{n^2}}$, ${\overline \Delta _M}\left( t \right) \in \mathbb{R}^{{n^2} \times{n^2}}$.

The equation \eqref{eqA1} is multiplied by $ adj\left\{ {{{\overline \Delta }_M}\left( t \right)} \right\}$: 
\begin{equation}
\begin{gathered}
vec\left( {{Y_M}\left( t \right)} \right){\rm{:}} = {\Delta _M}\left( t \right)vec\left( M \right){\rm{,}}
\end{gathered}\normalsize{\label{eqA2}}
\end{equation}
where ${Y_M}\left( t \right) \in \mathbb{R}^{n \times n}$, ${\Delta _M}\left( t \right){\rm{:}} = det\left\{ {{{\overline \Delta }_M}\left( t \right)} \right\} \in \mathbb{R}$.

The operation $ve{c^{ - 1}}\left( . \right)$ is applied to \eqref{eqA2}:
\begin{equation}
\begin{gathered}
	{Y_M}\left( t \right) = {\Delta _M}\left( t \right)M{\rm{.}}
	\end{gathered}\normalsize{\label{eqA3}}
\end{equation}

The $2^{nd}$ equation of \eqref{eq8} is transposed and multiplied by ${\Delta _M}\left( t \right)$:
\begin{equation}
\begin{gathered}
{\overline Y_x}\left( t \right){\rm{:}} = {\Delta _M}\left( t \right)h = {\overline \Delta _x}\left( t \right)K_x^{\rm{T}}{\rm{,}}
\end{gathered}\normalsize{\label{eqA4}}
\end{equation}
where ${\overline Y_x}\left( t \right) \in \mathbb{R}^n$, ${\overline \Delta _x}{\rm{:}} = Y_M^{\rm{T}} \in \mathbb{R} ^{n \times n}$.

Having multiplied \eqref{eqA4} by $adj\left\{ {{{\overline \Delta }_x}\left( t \right)} \right\}$, it is obtained:
\textcolor{black}{\begin{equation}
\begin{gathered}
{Y_x}\left( t \right) : = \left( adj\left\{ {{{\overline \Delta }_x}\left( t \right)} \right\}{\Delta _M}\left( t \right)h\right)^{\rm{T}} = {\Delta _x}\left( t \right){K_x}{\rm{,}}
\end{gathered}\normalsize{\label{eqA5}}
\end{equation}
where ${Y_x}\left( t \right) \in \mathbb{R}^{1 \times n}$, ${\Delta _x}\left( t \right) = det\left\{ {{{\overline \Delta }_x}\left( t \right)} \right\} \in \mathbb{R}$.}

\textcolor{black}{Then, the next aim is to obtain the regression with respect to $K_{r}$.} The equation \eqref{eq10} is multiplied by $ \varphi \left( t \right)$, and $z_{B}\left( t \right)$  from \eqref{eq17} is substituted into the obtained result:
\begin{equation}
\begin{gathered}
- {h^{\rm{T}}}A_\Sigma ^{ - 1}{z_B}\left( t \right){K_r} = \varphi \left( t \right){\rm{,}}
\end{gathered}\normalsize{\label{eqA6}}
\end{equation}

In \eqref{eqA6} ${h^{\rm{T}}}A_\Sigma ^{ - 1}$ is unmeasurable since $A_\Sigma ^{ - 1}$ is unknown. However, ${h^{\rm{T}}}A_\Sigma ^{ - 1}$ could be estimated with the help of \eqref{eq8} and \eqref{eqA3}, \eqref{eqA5}:
\begin{equation}
\begin{gathered}
M\Gamma  = {A_\Sigma }M \Rightarrow \left. {{\Gamma ^{ - 1}}{M^{ - 1}} = {M^{ - 1}}A_\Sigma ^{ - 1}} \right| \times M, \Rightarrow\\
\begin{array}{c}
\left. {M{\Gamma ^{ - 1}}{M^{ - 1}} = A_\Sigma ^{ - 1}} \right| \times {h^{\rm{T}}} \Rightarrow \\
\left. {{h^{\rm{T}}}M{\Gamma ^{ - 1}}{M^{ - 1}} = {h^{\rm{T}}}A_\Sigma ^{ - 1}} \right| \times {\Delta _M}\left( t \right) \Rightarrow \\
 \Rightarrow {h^{\rm{T}}}{Y_M}\left( t \right){\Gamma ^{ - 1}}{M^{ - 1}} = {h^{\rm{T}}}A_\Sigma ^{ - 1}{\Delta _M}\left( t \right).
\end{array}
\end{gathered}\normalsize{\label{eqA7}}
\end{equation}

In \eqref{eqA7} ${M^{ - 1}}$ is unknown, so let the regression with respect to ${M^{ - 1}}$ be obtained, for which \eqref{eqA3} is multiplied by ${M^{ - 1}}$:
\begin{equation}
\begin{gathered}
{\Delta _M}\left( t \right) = {Y_M}\left( t \right){M^{ - 1}}{\rm{,}}
\end{gathered}\normalsize{\label{eqA8}}
\end{equation}

The equation \eqref{eqA8} is multiplied by $adj\left\{ {{Y_M}\left( t \right)} \right\}$:
 \begin{equation}
\begin{gathered}
 {Y_{{M^{ - 1}}}}\left( t \right){\rm{:}} = adj\left\{ {{Y_M}\left( t \right)} \right\}{\Delta _M}\left( t \right) = {\Delta _{{M^{ - 1}}}}\left( t \right){M^{ - 1}}{\rm{,}}
 \end{gathered}\normalsize{\label{eqA9}}
\end{equation}
where ${Y_{{M^{ - 1}}}}\left( t \right) \in \mathbb{R} ^{n \times n}$,
${\Delta _{{M^{ - 1}}}}\left( t \right){\rm{:}} = det\left\{ {{Y_M}\left( t \right)} \right\} \in \mathbb{R}$.

Then \eqref{eqA7} is multiplied by ${\Delta _{{M^{ - 1}}}}\left( t \right)$, and \eqref{eqA9} is substituted into the result to obtain:
 \begin{equation}
\begin{gathered}
\begin{array}{c}
\left. {{h^{\rm{T}}}{Y_M}{\Gamma ^{ - 1}}{M^{ - 1}} = {h^{\rm{T}}}A_\Sigma ^{ - 1}{\Delta _M}} \right| \times {\Delta _{{M^{ - 1}}}} \Rightarrow \\
 \Rightarrow {h^{\rm{T}}}{Y_M}{\Gamma ^{ - 1}}{Y_{{M^{ - 1}}}} = {h^{\rm{T}}}A_\Sigma ^{ - 1}{\Delta _M}{\Delta _{{M^{ - 1}}}}.
\end{array}
 \end{gathered}\normalsize{\label{eqA10}}
\end{equation}

According to \eqref{eqA10}, $ {h^{\rm{T}}}A_\Sigma ^{ - 1}{\Delta _M}\left( t \right){\Delta _{{M^{ - 1}}}}\left( t \right)$ is known. So, having multiplied \eqref{eqA6} by ${\Delta _M}\left( t \right){\Delta _{{M^{ - 1}}}}\left( t \right)$, it is obtained:
 \begin{equation}
\begin{gathered}
{Y_r}\left( t \right){\rm{:}} =  - \varphi \left( t \right){\Delta _M}\left( t \right){\Delta _{{M^{ - 1}}}}\left( t \right) = {\Delta _r}\left( t \right){K_r}{\rm{,}}
\end{gathered}\normalsize{\label{eqA11}}
\end{equation}
where ${Y_r}\left( t \right) \in \mathbb{R}$, $ {\Delta _r}\left( t \right){\rm{:}} = {h^{\rm{T}}}{Y_M}{\Gamma ^{ - 1}}{Y_{{M^{ - 1}}}}{z_B} \in \mathbb{R}$.

Equations \eqref{eqA5} and \eqref{eqA11} are rewritten in the matrix form:
\begin{equation}
\small
\begin{gathered}
\overline Y\left( t \right) = \overline \Delta \left( t \right)\theta {\rm{,}}\\ \overline Y\left( t \right){\rm{:}} = 
\begin{bmatrix}
{{Y_x}\left( t \right)}\!\!\!\! & {{Y_r}\left( t \right)}
\end{bmatrix}^{\rm{T}}\!\!,\:\overline \Delta \left( t \right){\rm{:}} \!=\! 
\bigg(\begin{smallmatrix}
{{\Delta _x}\left( t \right)}{I_{n \times n}}&0_n\\
0_{1 \times n}&{{\Delta _r}\left( t \right)}\\
\end{smallmatrix}\bigg) ,
\end{gathered}\normalsize{\label{eqA12}}
\end{equation}
where $\overline Y\left( t \right){ \in \mathbb{R}^{n + 1}}$, $\overline \Delta \left( t \right){ \in \mathbb{R}^{\left( {n + 1} \right) \times \left( {n + 1} \right)}}$.

Having multiplied \eqref{eqA12} by $adj\left\{ {\overline \Delta \left( t \right)} \right\}$, according to \eqref{eqA1}-\eqref{eqA11}, the equation \eqref{eq18} is obtained exactly to the notation:
\begin{equation}
\begin{gathered}
\begin{array}{c}
Y{\rm{ }}\left( t \right){\rm{:}} = adj\left\{ {\overline \Delta \left( t \right)} \right\}\overline Y\left( t \right){\rm{, }}\\\Delta {\rm{ }}\left( t \right){\rm{:}} = \Delta _x^n\left( t \right){\Delta _r}\left( t \right) = C{\varphi ^{q}}\left( t \right),\\
C{\rm{:}} = C_1^{{n^2} + n + 1}C_2^{n + 1}{C_3}, q : = {n^4} + {n^3} + {n^2} + 1,
\\
{\rm{ }}{C_1}{\rm{:}} = det\left\{ { - {I_{n \times n}} \otimes A + {\Gamma ^{\rm{T}}} \otimes {I_{n \times n}}} \right\}{\rm{, }}\\{C_2}{\rm{:}} = det\left\{ M \right\}{\rm{, }} \ {C_3}{\rm{:}} = {h^{\rm{T}}}A_\Sigma ^{ - 1}B,
\end{array}
\end{gathered}\normalsize{\label{eqA13}}
\end{equation}
where $Y\left( t \right){ \in \mathbb{R}^{n + 1}}$ is measurable because ${\overline \Delta \left( t \right)}$ and $\overline Y\left( t \right)$ are measurable according to \eqref{eq17}, \eqref{eqA1}-\eqref{eqA11}, $\Delta \left( t \right) \in \mathbb{R}$ is measurable as $\Delta _{x}\left( t \right)$ and $\Delta_{r}\left( t \right)$ are measurable. \textcolor{black}{More details on how to obtain \eqref{eqA13} could be found in \cite{b33} (Section II).}

\textcolor{black}{Then the next aim is to show that $\Delta \left( t \right) \in {\rm{FE}}$.} According to Remark 2, when $\Phi \left( t \right) \in {\rm{FE}}$, then $\varphi \left( t \right) \in {\rm{FE}}$ holds. So, the implication $\varphi \left( t \right) \in {\rm{FE}} \Rightarrow \Delta \left( t \right) \in {\rm{FE}}$ is the only one, which needs to be proved. For this purpose, the condition \eqref{eq2} is written for $ \varphi \left( t \right)$:
\begin{equation}
\begin{gathered}
\small\int \limits_{t_r^ + }^{{t_e}} {{\varphi ^2}\left( \tau  \right)d\tau }  \ge \alpha .
\end{gathered}\normalsize{\label{eqA14}}
\end{equation}

It is easy to show that \eqref{eqA14} holds if $\exists {t_0} \in \left[ {{t_r^ + }{\rm{; }}t_e } \right]$ such that ${\rm{ }}{\varphi ^2}\left( {{t_0}} \right) > 0$. In fact, let ${\varphi ^2}\left( {{t_0}} \right) = \beta  > 0$, then, as ${\varphi ^2}\left( t \right)$ is continuous, such a neighborhood of $t_{0}$ exists that the inequality ${\varphi ^2}\left( t \right) \ge {\textstyle{\beta  \over 2}}$ holds for each segment $\left[ {{t_a}{\rm{; }}{t_b}} \right]$ of such neighborhood. Then the following holds:
\begin{equation}
\begin{gathered}
\small\int \limits_{t_r^ + }^{{t_e}} {{\varphi ^2}\left( \tau  \right)d\tau } \ge \int\limits_{t_r^ + }^{{t_a}} {{\varphi ^2}\left( \tau  \right)d\tau }  + \int\limits_{{t_a}}^{{t_b}} {{\varphi ^2}\left( \tau  \right)d\tau } \\
+ \int\limits_{{t_b}}^{{t_e}} {{\varphi ^2}\left( \tau  \right)d\tau } \ge \int\limits_{{t_a}}^{{t_b}} {{\varphi ^2}\left( \tau  \right)d\tau }  \ge {\textstyle{\beta  \over 2}}\small\int \limits_{{t_a}}^{{t_b}} {1d\tau }   =\\
={\textstyle{\beta  \over 2}}\left( {{t_b} - {t_a}} \right) = \alpha {\rm{,}}
\end{gathered}\normalsize{\label{eqA15}}
\end{equation}

Then, when $\varphi \left( t \right) \in {\rm{FE}}$, the inequality ${\varphi ^{2q}}\left( t \right) \ge {\textstyle{{{\beta ^{q}}} \over {{2^{q}}}}}$ holds for a segment $\left[ {{t_a}{\rm{; }}{t_b}} \right]$. So, the condition \eqref{eq2} for $\Delta \left( t \right)$ is rewritten as:
\begin{equation}
\begin{gathered}
\small\int\limits_{t_r^ + }^{{t_e}} {{\Delta ^2}\left( \tau  \right)d\tau } =\int\limits_{t_r^ + }^{{t_a}} {{\Delta ^2}\left( \tau  \right)d\tau }  + \int\limits_{{t_a}}^{{t_b}} {{\Delta ^2}\left( \tau  \right)d\tau }  +\\
+\int\limits_{{t_b}}^{{t_e}} {{\Delta ^2}\left( \tau  \right)d\tau }  \ge \int\limits_{{t_a}}^{{t_b}} {{\Delta ^2}\left( \tau  \right)d\tau }  \ge \\
\ge {C^2}\left( {{t_b} - {t_a}} \right){\left( {{\textstyle{\alpha  \over {{t_b} - {t_a}}}}} \right)^{2q}}{\rm{,}}
\end{gathered}\normalsize{\label{eqA16}}
\end{equation}

According to Proposition 1, first of all, $\exists {M^{ - 1}}$. This means that ${C_2} \ne 0$ and $\left( { - {I_{n}} \otimes {z_A} + \varphi {\Gamma ^{\rm{T}}}} \right)vec\left( M \right) = vec\left( {B{h^{\rm{T}}}} \right)$ is solvable. In its turn, this means that ${C_1} \ne 0$. As, according to \eqref{eq10}, ${h^{\rm{T}}}A_\Sigma ^{ - 1}B = K_r^{ - 1} = {C_3} \ne 0$, then $C = const \ne 0$. So, using \eqref{eqA13} and \eqref{eq2}, it follows that $\varphi \left( t \right) \in {\rm{FE}} \Rightarrow \Delta \left( t \right) \in {\rm{FE}}$.

\end{document}